\newcommand{\CLASSINPUTtoptextmargin}{1in}
\newcommand{\CLASSINPUTbottomtextmargin}{0.75in}

\documentclass[conference]{IEEEtran}
\IEEEoverridecommandlockouts
\usepackage{amsmath,amssymb,amsfonts}
\usepackage{graphicx}
\usepackage[caption=false, font=footnotesize]{subfig}
\usepackage{textcomp}
\usepackage{threeparttable}
\usepackage{multirow}
\usepackage{adjustbox}
\usepackage[table,xcdraw]{xcolor}
\usepackage{svg}
\usepackage{calc}
\usepackage[acronym]{glossaries}
\usepackage{comment}

\usepackage{nohyperref}

\definecolor{matplotlib0}{HTML}{1f77b4}
\definecolor{matplotlib1}{HTML}{d62728}
\definecolor{matplotlib2}{HTML}{2ca02c}
\definecolor{matplotlib3}{HTML}{ff7f0e}
\definecolor{matplotlib4}{HTML}{9467bd}
\definecolor{matplotlib5}{HTML}{8c564b}
\definecolor{matplotlib6}{HTML}{e377c2}
\definecolor{matplotlib7}{HTML}{7f7f7f}
\definecolor{matplotlib8}{HTML}{bcbd22}
\definecolor{matplotlib9}{HTML}{17becf}

\usepackage{mathtools} 

\usepackage{booktabs}
\usepackage{multirow}
\usepackage{colortbl}
\usepackage{tablefootnote}
\usepackage{threeparttable}

\usepackage{tikz}

\usepackage{pgfplots}
\definecolor{color0}{rgb}{0.12156862745098,0.466666666666667,0.705882352941177} 
\definecolor{color1}{rgb}{1,0.498039215686275,0.0549019607843137}
\definecolor{color2}{rgb}{0.172549019607843,0.627450980392157,0.172549019607843} 
\definecolor{color3}{rgb}{0.83921568627451,0.152941176470588,0.156862745098039} 
\definecolor{color4}{rgb}{0.580392156862745,0.403921568627451,0.741176470588235}
\definecolor{colorblue}{rgb}{0.12156862745098,0.466666666666667,0.705882352941177} 
\definecolor{colorgreen}{rgb}{0.172549019607843,0.627450980392157,0.172549019607843} 
\definecolor{colorred}{rgb}{0.83921568627451,0.152941176470588,0.156862745098039} 
\definecolor{colorblack}{rgb}{0,0,0} 
\definecolor{amber}{rgb}{1.0, 0.75, 0.0}
\definecolor{cyan(process)}{rgb}{0.0, 0.72, 0.92}
\definecolor{colororange}{rgb}{1,0.56,0} 
\definecolor{colorbrown}{rgb}{0.62, 0.42, 0.21} 
\definecolor{colorpurple}{rgb}{0.33, 0.033, 0.51}
\usepgfplotslibrary{fillbetween}
\usepgfplotslibrary{colormaps}
\pgfplotsset{compat=1.16}

\pgfplotscreateplotcyclelist{matplotlib}{
  {matplotlib0},
  {matplotlib1},
  {matplotlib2},
  {matplotlib3},
  {matplotlib4},
  {matplotlib5},
  {matplotlib6},
  {matplotlib7},
  {matplotlib8},
  {matplotlib9}
}
\pgfplotsset{every axis/.append style={
    cycle list name=matplotlib,
}}
\pgfdeclareplotmark{mystar}{
    \node[star,star point ratio=2.25,minimum size=6pt,
          inner sep=0pt,draw=black,solid,fill=red] {};
}

\usepackage{listings}
\usepgfplotslibrary{groupplots}

\definecolor{code_default}{HTML}{000000}
\definecolor{code_keyword}{HTML}{AC4142}
\definecolor{code_identifier}{HTML}{D28445}

\lstdefinelanguage{RISCV}{
  sensitive=false,
  morecomment=[l]{//},
  alsoletter={.},
  morekeywords=[1]{
    lp.setup, mv, lw, p.lw, sw, p.sw, pv.sdotsp.b, pv.shuffle2.b, p.subNR, p.addNR
  },
  morekeywords=[2]{
    zero, ra, sp, gp, tp, t0, t1, t2, t3, t4, t5, t6, s0, s1, a0, a1, a2, a3, a4, a5, a6, a7, a8, a9, a10, a11,
  },
  morestring=[b]",
  morestring=[b]',
}[strings, comments, keywords]

\lstdefinestyle{RISCV_STYLE}{
  language=RISCV,
  numbers=none,
  basicstyle=\scriptsize\ttfamily\color{code_default},
  keywordstyle=[1]\color{matplotlib0},
  keywordstyle=[2]\color{matplotlib1},
  float,
  captionpos=b,
  belowskip=-0.5cm
}

\usepackage{algorithm}
\usepackage{algpseudocode}
\usepackage{float}
\newfloat{algorithm}{t}{top}

\renewcommand{\baselinestretch}{0.97}
\def\BibTeX{{\rm B\kern-.05em{\sc i\kern-.025em b}\kern-.08em
    T\kern-.1667em\lower.7ex\hbox{E}\kern-.125emX}}

\newcommand{\seizurenetwork}[0]{\textsc{EpiDeNet}}

\makeatletter
\def\ps@IEEEtitlepagestyle{%
  \def\@oddfoot{\mycopyrightnotice}%
  \def\@oddhead{\hbox{}\@IEEEheaderstyle\leftmark\hfil\thepage}\relax
  \def\@evenhead{\@IEEEheaderstyle\thepage\hfil\leftmark\hbox{}}\relax
  \def\@evenfoot{}%
}
\def\mycopyrightnotice{%
  \begin{minipage}{\textwidth}
  \centering \scriptsize
  \copyright 2023 IEEE.  Personal use of this material is permitted.  Permission from IEEE must be obtained for all other uses, in any current or future media, including reprinting/republishing this material for advertising or promotional purposes, creating new collective works, for resale or redistribution to servers or lists, or reuse of any copyrighted component of this work in other works.
  \end{minipage}
}
\makeatother
\begin{document}

\title{\seizurenetwork: An Energy-Efficient Approach to Seizure Detection for Embedded Systems}
 \author{\IEEEauthorblockN{
    Thorir~Mar~Ingolfsson\IEEEauthorrefmark{1},
    Upasana~Chakraborty\IEEEauthorrefmark{1},
    Xiaying~Wang\IEEEauthorrefmark{1}, 
    Sandor~Beniczky\IEEEauthorrefmark{4},\IEEEauthorrefmark{6}\\
    Pauline~Ducouret\IEEEauthorrefmark{3},
    Simone~Benatti\IEEEauthorrefmark{5},
    Philippe~Ryvlin\IEEEauthorrefmark{3},
    Andrea~Cossettini\IEEEauthorrefmark{1},
    Luca~Benini\IEEEauthorrefmark{1}\IEEEauthorrefmark{2}}
     
    \vspace{0.2cm}

    \IEEEauthorblockA{\\\IEEEauthorrefmark{1}ETH Z{\"u}rich, Z{\"u}rich, Switzerland \hspace{3.2mm}\IEEEauthorrefmark{2}University of Bologna, Bologna, Italy \hspace{3.2mm}\IEEEauthorrefmark{4}Aarhus University Hospital, Aarhus, Denmark}
    \IEEEauthorblockA{\IEEEauthorrefmark{5}Università di Modena e Reggio Emilia, Reggio Emilia, Italy \hspace{3.2mm}\IEEEauthorrefmark{3}Lausanne  University  Hospital (CHUV), Switzerland}
    \IEEEauthorblockA{\IEEEauthorrefmark{6}Danish Epilepsy Centre (Filadelfia), Dianalund, Denmark\vspace{-0.2cm}}
    
    \thanks{Corresponding email: \{thoriri\}@iis.ee.ethz.ch}
    
    \vspace{-0.5cm}
    }
\maketitle
\newacronym{ofa}{OFA}{Once-For-All}
\newacronym{simd}{SIMD}{Single Instruction, Multiple Data}
\newacronym{elu}{ELU}{Exponential Linear Unit}
\newacronym{relu}{ReLU}{Rectified Linear Unit}
\newacronym{rpr}{RPR}{Random Partition Relaxation}
\newacronym{mac}{MAC}{Multiply Accumulate}
\newacronym{dma}{DMA}{Direct Memory Access}
\newacronym{bmi}{BMI}{Brain--Machine Interface}
\newacronym{bci}{BCI}{Brain--Computer Interface}
\newacronym{smr}{SMR}{Sensory Motor Rythms}
\newacronym{eeg}{EEG}{Electroencephalography}
\newacronym{svm}{SVM}{Support Vector Machine}
\newacronym{svd}{SVD}{Singular Value Decomposition}
\newacronym{evd}{EVD}{Eigendecomposition}
\newacronym{iir}{IIR}{Infinite Impulse Response}
\newacronym{fir}{FIR}{Finite Impulse Response}
\newacronym{fc}{FC}{Fabric Controller}
\newacronym{nn}{NN}{Neural Network}
\newacronym{mrc}{MRC}{Multiscale Riemannian Classifier}
\newacronym{flop}{FLOP}{Floating Point Operation}
\newacronym{sos}{SOS}{Second-Order Section}
\newacronym{ipc}{IPC}{Instructions per Cycle}
\newacronym{tcdm}{TCDM}{Tightly Coupled Data Memory}
\newacronym{fpu}{FPU}{Floating Point Unit}
\newacronym{fma}{FMA}{Fused Multiply Add}
\newacronym{alu}{ALU}{Arithmetic Logic Unit}
\newacronym{dsp}{DSP}{Digital Signal Processing}
\newacronym{gpu}{GPU}{Graphics Processing Unit}
\newacronym{soc}{SoC}{System-on-Chip}
\newacronym{mi}{MI}{Motor-Imagery}
\newacronym{csp}{CSP}{Commmon Spatial Patterns}
\newacronym{fbcsp}{FBCSP}{Filter-Bank \acrlong{csp}}
\newacronym{pulp}{PULP}{parallel ultra-low power}
\newacronym{soa}{SoA}{state-of-the-art}
\newacronym{bn}{BN}{Batch Normalization}
\newacronym{isa}{ISA}{Instruction Set Architecture}
\newacronym{ecg}{ECG}{Electrocardiogram}
\newacronym{mcu}{MCU}{microcontroller}
\newacronym{rnn}{RNN}{recurrent neural network}
\newacronym{cnn}{CNN}{convolutional neural network}
\newacronym{tcn}{TCN}{temporal convolutional network}
\newacronym{emu}{EMU}{epilepsy monitoring unit}
\newacronym{ml}{ML}{Machine Learning}
\newacronym{dl}{DL}{Deep Learning}
\newacronym{ai}{AI}{Artificial Intelligence}
\newacronym{tcp}{TCP}{Temporal Central Parasagittal}
\newacronym{loocv}{LOOCV}{Leave-One-Out Cross-Validation}
\newacronym{wfcv}{WFCV}{Walk-Forward Cross-Validation}
\newacronym{rwcv}{RWCV}{Rolling Window Cross-Validation}
\newacronym{iot}{IoT}{Internet of Things}
\newacronym{auc}{AUC}{Area Under the Receiver Operator Characteristic}
\newacronym{dwt}{DWT}{Discrete Wavelet Transform}
\newacronym{fft}{FFT}{Fast Fourier Transform}
\newacronym{tpot}{TPOT}{Tree-based Pipeline Optimization Tool}

\newacronym{tuar}{TUAR}{Temple University Artifact Corpus}
\newacronym{tuev}{TUEV}{Temple University Event Corpus}

\newacronym{bss}{BSS}{Blind Source Separation}
\newacronym{ica}{ICA}{Independent Component Analysis}
\newacronym{ic}{ICs}{Independent Components}
\newacronym{asr}{ASR}{Artifact Subspace Reconstruction}
\newacronym{pca}{PCA}{Principal Component Analysis}
\newacronym{gap}{GAP}{Global Average Pooling}
\newacronym{fcn}{FCN}{Fully Connected Networks}
\newacronym{mlp}{MLP}{Multi-Layer Perceptron}
\newacronym{nas}{NAS}{Neural Architectural Search}
\newacronym{fph}{FP/h}{False Positives per Hour}
\newacronym{bvp}{BVP}{Blood volume Pulse}
\newacronym{eda}{EDA}{Electrodermal Activity}
\newacronym{acc}{ACC}{Accelerometer}
\newacronym{cae}{CAE}{Convolutional Autoencoder}
\newacronym{sswce}{SSWCE}{Sensitivity-Specificity Weighted Cross-Entropy}
\newacronym{ce}{CE}{Cross-Entropy}

\begin{abstract}
Epilepsy is a prevalent neurological disorder that affects millions of individuals globally, and continuous monitoring coupled with automated seizure detection appears as a necessity for effective patient treatment. To enable long-term care in daily-life conditions, comfortable and smart wearable devices with long battery life are required, which in turn set the demand for resource-constrained and energy-efficient computing solutions. In this context, the development of machine learning algorithms for seizure detection faces the challenge of heavily imbalanced datasets. This paper introduces \seizurenetwork, a new lightweight seizure detection network, and \gls{sswce}, a new loss function that incorporates sensitivity and specificity, to address the challenge of heavily unbalanced datasets. The proposed \seizurenetwork-\gls{sswce} approach demonstrates the successful detection of $91.16\%$ and $92.00\%$ seizure events on two different datasets (CHB-MIT and PEDESITE, respectively), with only four EEG channels.
A three-window majority voting-based smoothing scheme combined with the \gls{sswce} loss achieves $3\times$ reduction of false positives to $1.18$ FP/h. \seizurenetwork{} is well suited for implementation on low-power embedded platforms, and we evaluate its performance on two ARM Cortex-based platforms (M4F/M7) and two \gls{pulp} systems (GAP8, GAP9). The most efficient implementation (GAP9) achieves an energy efficiency of $40$ GMAC/s/W, with an energy consumption per inference of only $0.051$ mJ at high performance ($726.46$ MMAC/s), outperforming the best ARM Cortex-based solutions by approximately $160\times$ in energy efficiency. The \seizurenetwork-\gls{sswce} method demonstrates effective and accurate seizure detection performance on heavily imbalanced datasets, while being suited for implementation on energy-constrained platforms.

\end{abstract}


\section{Introduction}
Epilepsy is a widespread neurological disorder affecting over 50 million individuals globally and is characterized by recurrent seizures that temporarily impair brain function~\cite{world2019epilepsy}. Traditional interventions involve pharmacological approaches; however, drug-resistant patients may require surgical procedures or invasive neurostimulation, necessitating individualized treatment plans and protracted cerebral activity recording.

Patient monitoring typically occurs in \glspl{emu}, where individuals are observed using video monitoring systems and noninvasive \gls{eeg} caps with 32 channels. \gls{eeg} data enables the identification of various seizure types, thereby going beyond generalized convulsive seizures which only account for less than $20\%$ of all seizures (and are often detected through non-\gls{eeg} signals). While \gls{emu} monitoring is very useful in characterizing seizure type, methods to enable long-term recording in ambulatory patients are still needed to provide a more precise assessment of seizure frequency and seizure-triggered alarms.~\cite{beniczky2017standardized,elger2018diagnostic,ryvlin2013incidence}

This paper addresses the challenge of \gls{eeg}-based seizure detection for wearable devices. Traditional \gls{eeg} systems are cumbersome, uncomfortable, and stigmatizing. Therefore, caregivers and patients prefer wearable solutions to facilitate long-term continuous \gls{eeg} monitoring. Moreover, integrating seizure detection capabilities through \gls{ml} into wearables is particularly interesting since it would enable timely interventions by caregivers during or immediately following seizures, thus reducing their impact and providing reliable information to physicians for optimizing anti-seizure therapies. However, several challenges still persist in developing effective seizure detection models for wearable devices.

First, many existing \gls{ai} models rely heavily on a large number of electrodes. While state-of-the-art performance with sensitivity and specificity greater than $98\%$ have been proposed \cite{abdel}, such solutions are not applicable to wearables for continuous monitoring, where the number of channels is minimal. 

Second, false alarms in long-term monitoring settings have a more pronounced impact on patients' and caregivers' willingness to use the devices. While a great majority of existing works focus on sensitivity, it is crucial to prioritize maximizing the specificity, even at the cost of not detecting some seizures, as the primary performance metric~\cite{bruno2018wearable}.

Third, the inherent variability in patient characteristics demands personalized approaches, as seizures can greatly vary due to seizure type, etiology, and unique brain characteristics~\cite{fisher2005epileptic, kuhlmann2018seizure}. Developing and optimizing subject-specific models is essential to cater to each patient's distinct needs and patterns~\cite{greene2007combination, truong2018convolutional}, enhancing seizure detection accuracy, improving wearable device performance, and providing more precise information for better therapeutic outcomes~\cite{cho2017robust}. 

Finally, wearable devices must meet several key criteria, such as a compact form factor, extended battery life, and minimal latency. Recent advancements have addressed these requirements by incorporating smart edge computing with low-power \glspl{mcu}, which has demonstrated efficacy in facilitating long-term operations and executing AI models~\cite{ingolfsson2021ecg}. Nevertheless, ensuring that AI algorithms align with the computational capacities of wearable devices is a notable challenge. Consequently, the selection of models must be refined for implementation in low-power \glspl{mcu}.

In this context, the study of~\cite{zhao2021energy} emphasized energy efficiency, wherein the authors employed \gls{nas} to design a CNN with approximately 11k model parameters. The estimated energy consumption for their model was approximately $3.26$ mJ per inference on an embedded platform. Although their CNN architecture achieved a reported sensitivity of $99.81\%$, it was not validated on an embedded device or designed to accommodate a lower channel count.  In contrast, \cite{eeg-transformer} represents the current benchmark, addressing the challenges associated with reduced electrode montage for wearable technologies and achieving $99.9\%$ specificity, $65.5\%$ sensitivity, and $0.8$ FP/h on the CHB-MIT dataset. However, this network was only validated on a subset of the CHB-MIT dataset, obscuring its generalizability across all patients or diverse datasets.


The main contributions of the paper are as follows:
\begin{itemize}
    \item Development and validation of \seizurenetwork{}, a lightweight and resource-friendly seizure detection network detecting $91.16\%$ and $92.00\%$ of seizures on the CHB-MIT and on the PEDESITE epilepsy datasets, respectively, only using temporal channels, while having $2.5$ FP/h.
    \item Introduction of a new loss function (\gls{sswce} loss) enhances the network's performance in seizure detection, leading to a $3.64\%$ increase in sensitivity and reducing the false alarms to only $1.18$ FP/h.
    \item Quantization and deployment of the trained model on ARM Cortex M4F/M7, RISC-V \gls{pulp}-based GreenWaves Technologies GAP8, and GAP9 MCUs using TFLite, CUBE.AI, and Quantlab/DORYtools~\cite{quantlab,burrello2020dory}, achieving state of the art energy efficiency and demonstrating the suitability of \seizurenetwork{} for wearable devices, with the GAP9 implementation achieving $40$ GMAC/s/W energy efficiency and the lowest energy consumption per inference ($0.051$ mJ).
\end{itemize}

\section{Methods}
\subsection{Model Architecture}
Table~\ref{tab:CNN_Table} presents the proposed model architecture, which is a refined version of the networks identified using the \gls{nas} algorithm, as described in~\cite{zhao2021energy}. The network is designed to learn and extract meaningful features from input data progressively. In the initial layers ($\phi^1$ to $\phi^3$), the network focuses on learning frequency filters, which are essential for capturing the spectral characteristics of the data. Subsequently, the intermediate layers ($\phi^4$–$\phi^5$) emphasize the acquisition of frequency-specific spatial filters, enabling the model to recognize spatial patterns contingent on specific frequency bands.

In the final block of the network ($\phi^6$), the architecture is engineered to integrate the feature maps produced by the preceding layers optimally. This approach facilitates the creation of a comprehensive and discriminative representation of the input data, bolstering the capacity of the model to classify and predict outcomes in complex scenarios accurately. We call this network \seizurenetwork{}.

\subsection{Sensitivity-Specificity Weighted Loss Function}
\Gls{ce} loss is a prevalent loss function for machine and deep learning classification problems. However, it can be unsuitable for imbalanced datasets, frequently encountered in epilepsy research, due to skewed class distributions, resulting in biased model predictions.

Alternative loss functions and techniques to address this issue include Focal Loss~\cite{lin2017focal}, Weighted \gls{ce} Loss~\cite{wang2016training}, and Oversampling and Undersampling techniques~\cite{chawla2002smote, drummond2003c4}. To overcome \gls{ce} loss limitations, we introduce domain-specific knowledge as defined in Equation~\eqref{eq:MC}, where $SN$ and $SP$ represent sensitivity and specificity, respectively.
\begin{equation}\label{eq:MC}
SSWCE(y, p) = CE(y,p) + \alpha (1-SP) + \beta (1-SN)
\end{equation}
\noindent, where $y$ is the true class label (0 or 1), and $p$, is the predicted probability of the positive class (1). User-defined $\alpha$ and $\beta$ hyperparameters allow targeting higher specificity or sensitivity. Owing to the weighing of sensitivity and specificity, we call this loss Sensitivity-Specificity Weighted Cross-Entropy.

\begin{table}
\caption{\seizurenetwork{}-architecture}
\label{tab:CNN_Table}
\begin{threeparttable}
\centering

\begin{tabular}{lllll}
 & \textit{\textbf{Type}} & \textit{\textbf{\#Filters}} & \multicolumn{1}{c}{\textit{\textbf{Kernel}}} & \textit{\textbf{Output}} \\ \hline

\multirow{2}{*}{$\phi^1$}
 & \multicolumn{1}{|l}{Conv2D} & \multicolumn{1}{l}{$4$} & \multicolumn{1}{l}{($1$, $4$)} & \multirow{1}{*}{($4$,$C$,$T$)} \\
 
 & \multicolumn{1}{|l}{MaxPool} & \multicolumn{1}{l}{} & \multicolumn{1}{l}{(1, 8)} &($4$,$C$,$T//8$) \\ \hline
 \multirow{2}{*}{$\phi^2$} & \multicolumn{1}{|l}{Conv2D} & \multicolumn{1}{l}{$16$} & \multicolumn{1}{l}{($1$, $16$)} & \multirow{1}{*}{($16$,$C$,$T//8$)} \\
 & \multicolumn{1}{|l}{MaxPool} & \multicolumn{1}{l}{} & \multicolumn{1}{l}{(1, 4)} & \multicolumn{1}{l}{($16$,$C$,$T//32$)} \\ \hline
 \multirow{2}{*}{$\phi^3$} & \multicolumn{1}{|l}{Conv2D} & \multicolumn{1}{l}{$16$} & \multicolumn{1}{l}{($1$, $8$)} & \multirow{1}{*}{($16$,$C$,$T//32$)} \\
 & \multicolumn{1}{|l}{MaxPool} & \multicolumn{1}{l}{} & \multicolumn{1}{l}{($1$, $4$)} & \multirow{1}{*}{($16$,$C$,$T//128$)} \\ \hline
\multirow{2}{*}{$\phi^4$} & \multicolumn{1}{|l}{Conv2D} & \multicolumn{1}{l}{$16$} & \multicolumn{1}{l}{($16$, $1$)} & \multirow{1}{*}{($16$,$C$,$T//128$)} \\
 & \multicolumn{1}{|l}{MaxPool} & \multicolumn{1}{l}{} & \multicolumn{1}{l}{($4$, $1$)} & \multirow{1}{*}{($16$,$C//4$,$T//128$)} \\ \hline
 \multirow{2}{*}{$\phi^5$} & \multicolumn{1}{|l}{Conv2D} & \multicolumn{1}{l}{$16$} & \multicolumn{1}{l}{($8$, $1$)} & \multirow{1}{*}{($16$,$C//4$,$T//128$)} \\
 & \multicolumn{1}{|l}{AdaptiveAveragePool} & \multicolumn{1}{l}{} & \multicolumn{1}{l}{} & \multirow{1}{*}{($16$,$1$,$1$)} \\ \hline
\multirow{1}{*}{$\phi^6$} & \multicolumn{1}{|l}{Dense} & \multicolumn{1}{l}{} & \multicolumn{1}{l}{} & \multirow{1}{*}{2} \\
\end{tabular}
\begin{tablenotes}\footnotesize
\item[] $C$ = number of EEG channels, $T$ = number of time samples.
\vspace{-10pt}
\end{tablenotes}
\end{threeparttable}
\end{table}

As the \gls{sswce} loss function permits independent weighting of either sensitivity or specificity, we conduct studies involving a grid search for the optimal combination of $\alpha$ and $\beta$ parameters for each patient in the PEDESITE dataset.

\subsection{Model Validation}
In accordance with previous works~\cite{thorir_biocas} that have demonstrated the superiority of subject-specific approaches for seizure detection model training, we exclusively focus on subject-specific models. This implies that each model is trained solely on data from the corresponding subject. Furthermore, to enhance the robustness and accuracy of the metrics reported for each model, we employ a \gls{loocv} scheme for training and testing. More specifically, we train on all records containing seizures except for one, and validate using the excluded record. Additionally, considering different random seeds, every reported number represents the average of 5 repetition runs.

\subsection{Embedded Platforms}
We adapt \seizurenetwork{} for three embedded platforms to address wearable seizure detection applications and evaluate their execution and energy consumption.

\subsubsection{ARM Cortex}
We examine the B-L475E-IOT01A STM32L4 Discovery kit with an ARM Cortex M4F core and the STM32 Nucleo-144 development board with an ARM Cortex M7 core. The STM32L4 kit offers 1 MB flash memory, $128$ kB SRAM, and operates at 80 MHz, while the Nucleo-144 kit provides $1$ MB flash memory, $320$ kB SRAM, and functions at 216 MHz.

\subsubsection{GAP8}
The GAPuino features a RISC-V-based PULP GAP8 processor with a dual compute domain architecture, including a fabric controller, a cluster domain with eight cores, 512 kB of L2 memory, and 80 kB of L1 memory, operating at frequencies between 32 kHz and 250 MHz.

\subsubsection{GAP9}
The GAP9 processor features ten cores divided between a Fabric Controller and Cluster Cores, 128 kB L1 memory, 1.5 MB RAM, and cluster cores operating at frequencies up to 370 MHz, sharing four Floating-Point Units. It optimally balances energy efficiency and computational performance within a milliwatt power envelope.

\subsection{Hardware Deployment}
For deploying \seizurenetwork{} on ARM-based platforms, we leverage \textsc{Cube.AI}, which streamlines the automatic conversion of pre-trained neural networks and seamlessly integrates the optimized library into the project code. \textsc{Cube.AI} takes a quantized INT-8 TFLite model as input, automatically parses the network graph, and produces optimized code accordingly.

For the GAP8/GAP9 platforms, we rely on Quantlab~\cite{quantlab} to quantize \seizurenetwork{} into an INT-8 format and on DORY~\cite{burrello2020dory} for the subsequent deployment. DORY is a dedicated tool that autonomously generates C code for managing the two-level memory hierarchy (L1 and L2 memory) on PULP-based platforms. While DORY natively supports code generation for GAP8, we manually optimize the generated code to ensure compatibility with the GAP9 platform.

\subsection{Datasets}
We present the seizure detection results for both the CHB-MIT~\cite{goldberger_physiobank_2000} and PEDESITE datasets. The CHB-MIT public dataset comprises of EEG data from 23 pediatric and young adult patients (aged 1.5--22 years) with intractable seizures, collected at 256 samples per second and a 16-bit resolution with 10/20 system. We use a $4$ second window which equates to $1024$ time samples -- (T in Table~\ref{tab:CNN_Table}).

Recognizing the CHB-MIT dataset's limitations in real-world scenarios (pediatric patients, gaps in the EEG traces, nearly zero artifacts), we also evaluate our approach using the private PEDESITE dataset from Lausanne University Hospital (CHUV). Patients undergo routine clinical evaluations during the study at the \gls{emu}. Scalp-EEG signals are collected at 1024 samples per second and a 16-bit resolution with a 10/20 system. Monitoring durations range from 2 days to two weeks. We further decimate the signal to 256 samples per second, to be comparable to the sampling rate of CHB-MIT, and look at an $8$ second window which is $2048$ time samples.

\vspace{-0.1cm}
\section{Seizure Detection Results}
\begin{table}[t]
\renewcommand{\arraystretch}{1}
\setlength{\tabcolsep}{4.3pt}
  \centering
  \caption{\gls{loocv} Comparison of~\cite{zhao2021energy} (All channels) and \seizurenetwork{} (Temporal Channels) on the CHB-MIT dataset. Bold results indicate a three-window majority smoothing scheme.}\label{tab:chb-mit-results2}
\begin{tabular}{ccccc}
\toprule
     Network & Sens. [\%] & Spec. [\%] & FP/h &\# Detected\\ 
     & & & & Seizures \\
    \midrule
    \cite{zhao2021energy} \textbf{22 Ch.}    & 82.77 (\textbf{81.91}) & 99.72 (\textbf{99.86}) & 2.53 (\textbf{1.25}) & \textbf{177/181} \\ 
    \seizurenetwork{} \textbf{4 Ch.} & 68.56 (\textbf{68.73}) & 99.52 (\textbf{99.75}) & 4.29 (\textbf{2.24}) & \textbf{165/181} \\
\bottomrule
\vspace{-15pt}
\end{tabular}
\end{table}

\begin{table}[b]
\vspace{-0.3cm}
\renewcommand{\arraystretch}{1}
  \centering
  \caption{\gls{loocv} results of \seizurenetwork{} on the PEDESITE dataset. Bold results indicate a three-window majority smoothing scheme.}\label{tab:results_PEDESITE:new_best}
  {
    \footnotesize
    \begin{tabular}{@{}ccccc@{}}
      \toprule
        & Sens. [\%] & Spec. [\%] & FP/h &\# Detected\\ 
     & & & & Seizures \\
      \midrule
      \textbf{EEG}  & 58.02 (\textbf{57.17}) & 99.23 (\textbf{99.51}) & 3.69 (\textbf{2.50}) & \textbf{23/25}\\
    \textbf{SSWCE EEG}  & 61.66 (\textbf{60.66}) & 99.66 (\textbf{99.74}) & 1.51 (\textbf{1.18}) & \textbf{23/25} \\
      \bottomrule
    \end{tabular}
  }
\end{table}

\begin{table*}[ht!]
\renewcommand{\arraystretch}{0.95}
  \centering
  \caption{Comparison Between \seizurenetwork{} on GAP9, GAP8, ARM Cortex M4F/M7 and comparison to related works}\label{tab:results:summary}
  \vspace{-.2cm}
  {
    \footnotesize
    \begin{tabular}{lrrrrrr}
      \toprule
      Network & \multicolumn{4}{c}{\seizurenetwork{}} & \multicolumn{1}{c}{\cite{eeg-transformer}} & \multicolumn{1}{c}{\cite{liu2021edge}}\\
      \cmidrule(r){1-1} \cmidrule(r){2-5} \cmidrule(r){6-6} \cmidrule(r){7-7}
      Platform & \multicolumn{1}{c}{GAP9} & \multicolumn{1}{c}{GAP8} & \multicolumn{1}{c}{STM32L475} & \multicolumn{1}{c}{STM32F756} &  \multicolumn{1}{c}{Apollo 4} & \multicolumn{1}{c}{nRF52840}\\
      \multirow{2}{*}{MCU} & \multicolumn{1}{c}{1+9$\times$RISCY} & \multicolumn{1}{c}{1+8$\times$CV32E40P} & \multicolumn{1}{c}{1$\times$Cortex M4F} & \multicolumn{1}{c}{1$\times$Cortex M7} & \multicolumn{1}{c}{1$\times$Cortex M4F} & \multicolumn{1}{c}{1$\times$Cortex M4F}\\
       & \multicolumn{1}{c}{@240\,MHz} & \multicolumn{1}{c}{@100\,MHz}& \multicolumn{1}{c}{@80\,MHz} & \multicolumn{1}{c}{@216\,MHz} & \multicolumn{1}{c}{@96\,MHz} & \multicolumn{1}{c}{@64\,MHz}\\
       \cmidrule(r){2-2} \cmidrule(r){3-3} \cmidrule(r){4-4} \cmidrule(r){5-5} \cmidrule(r){6-6} \cmidrule(r){7-7}
      Deployment framework & \multicolumn{2}{c}{Quantlab/DORY} & \multicolumn{2}{c}{\textsc{Cube.AI} TFLite} & Quantlab/CMSIS-NN & TFLite \\ 
      Dataset & \multicolumn{4}{c}{PEDESITE} & \multicolumn{2}{c}{CHB-MIT} \\
     \cmidrule(r){1-1}\cmidrule(r){2-5} \cmidrule(r){6-7} 
      Input size & $4\times2048$ & $4\times2048$ & $4\times2048$  & $4\times2048$ & $4\times2048$ & $9\times256$\\
      MACs & 2\,064\,352 & 2\,064\,352  & 2\,053\,516 & 2\,053\,516  & 6\,250\,000 & 2\,400\,000 \\
      \cmidrule(r){1-1} \cmidrule(r){2-2}\cmidrule(r){3-3} \cmidrule(r){4-4} \cmidrule(r){5-5} \cmidrule(r){6-6} \cmidrule(r){7-7}
      Time/inference [ms]   & \textbf{2.84} & 6.82 &  190.60 & 31.99 &  405.00 & 100\\
      Throughput [MMAC/s] & \textbf{726.45} & 305.83 &  10.78 &  64.19 & 15.43 & 24 \\
      MACs/cycle & 3.03 & 3.03 &  0.135 & 0.30 & 0.16 & 0.38 \\
      Power [mW] & \textbf{17.89} &38.52 & 42.44 & 413.03 & 4.40 & 1.5\\
      Energy/inference [mJ]   & \textbf{0.051} &0.26& 8.09& 13.21 & 1.79 & 0.15\\
      En. eff. [GMAC/s/W] & \textbf{40.61} &7.86 & 0.25 & 0.15 & 3.51 &  16.00 \\
      \bottomrule
    \end{tabular}
  }
  
  \vspace{-.3cm}
\end{table*}
\subsection{CHB-MIT}
We validate the accuracy of the model proposed in~\cite{zhao2021energy} by re-implementing it and assessing its performance using the \gls{loocv} method. Table~\ref{tab:chb-mit-results2} shows the corresponding performance (first row). Notably, the model achieves a significantly lower sensitivity compared to the one reported in~\cite{zhao2021energy} ($82.77\%$ vs $99.81\%$). This discrepancy can be attributed to two factors: first, the authors of~\cite{zhao2021energy} do not consider the entire CHB-MIT dataset, but rather select 10 out of the 23 available patients; second, the rigorous method of testing and validating the models (leave-one-out) employed in this study is not utilized in~\cite{zhao2021energy}, potentially indicating that the results presented in~\cite{zhao2021energy} are overly optimistic.

Table~\ref{tab:chb-mit-results2} (second row) reports the performance of \seizurenetwork{}, built as an adaptation of~\cite{zhao2021energy} when considering only 4 channels in the temporal region (as common for low-power, non-stigmatizing wearable technologies). The authors of~\cite{zhao2021energy} do not report on the specificity of their network but reproduced numbers showcase that it achieves a very high specificity ($99.75\%$ - $99.86\%$). A noticeable decrease in sensitivity from ($81.91\% \rightarrow 68.73\%$) aligns with the figures documented in~\cite{eeg-transformer}, which similarly employs a four-channel temporal framework. It merits highlighting that the authors of \cite{eeg-transformer} incorporated only a subset of the total 23 patients (specifically 8), hence constraining a holistic comparison of sensitivity and specificity metrics. We reconstructed the~\cite{eeg-transformer} network to rectify this discrepancy and applied the identical train-test protocol to all CHB-MIT patients. The consequent specificity and sensitivity were reduced to $99.18\%$ and $59.91\%$, respectively, inflating false positives to $3.68$ FP/h. Compared with our \seizurenetwork{}, shows that \seizurenetwork{} excels with an elevated sensitivity by $8.8\%$ and a decline in false positives by $1.44$ per hour. Table~\ref{tab:chb-mit-results2} also reports the impact of employing a three-window majority voting-based smoothing scheme (bold numbers). This approach yields substantial reductions in false positives by approximately $2\times$, while maintaining a high seizure detection performance ($91\%$ of seizures are detected).

\subsection{PEDESITE}
We further validate \seizurenetwork{} on the PEDESITE dataset, as presented in Table~\ref{tab:results_PEDESITE:new_best}. The individual 8s EEG windows exhibit a relatively high number of false positives per hour, akin to the CHB-MIT dataset. Utilizing a three-window smoothing technique effectively reduces false positives to approximately $2.5$ per hour, with a negligible ($1\%$) decrement in sensitivity and still maintaining unaltered the high number of detected seizures ($23$ out of $25$, i.e., $92\%$).

To further improve the performance and showcase the impact of the proposed \gls{sswce} loss function, we optimize the $\alpha$ and $\beta$ parameters on a subject-specific basis, obtaining a $3.64\%$ increase in sensitivity and reducing the FP/h rate to $1.18$ (mostly caused by artifacts). To further minimize false positives, an artifact detector, such as the one proposed in~\cite{thorir_embc}, can be incorporated, as \gls{eeg} artifacts are susceptible to misclassification as seizures.

\subsection{Embedded Deployment}
We quantize \seizurenetwork{} using Quantlab for the GAP implementations and TFLite for the ARM implementations. In both cases, the change in accuracy is negligible, with approximately $\pm 0.1\%$ variations observed for both sensitivity and specificity.

After training the network, which relies on the \gls{sswce} loss function, we evaluate the energy/inference of the proposed \seizurenetwork{} network. To this extent we perform power measurements on four \glspl{mcu} and table~\ref{tab:results:summary} presents the comparison of measurements obtained from these methods and a comparison to two other works~\cite{eeg-transformer, liu2021edge}.

The GAP8 and GAP9 implementations outperform the ARM counterparts when utilizing Quantlab and DORY tools. Furthermore, DORY, supplemented with custom code to ensure smooth execution on GAP9, exhibits the lowest energy consumption per inference ($0.051$ mJ) and the highest energy efficiency ($40.61$ GMAC/s/W). This configuration also delivers the most substantial performance ($726.46$ MMAC/s), completing one inference in a mere $2.84$ ms. The most energy-efficient implementation (GAP9) is approximately $160\times$ more efficient than the optimal deployment of \seizurenetwork{} on the ARM Cortex MCUs. Comparing the implementation of \seizurenetwork{} to those in~\cite{eeg-transformer} and~\cite{liu2021edge}, the GAP9 implementation is $11.6\times$ and $2.6\times$ more energy-efficient, respectively.

\input{figures/power_trace}
\section{Conclusion}
\vspace{-0.2cm}
This work introduces \seizurenetwork{}, a compact model implemented on various low-power platforms, including GAP8, ARM Cortex M4F, ARM Cortex M7, and GAP9 processors. The model's energy consumption was as low as $0.051$ mJ per inference, with an average power consumption of $17.89$ mW on the GAP9 processor and more than two orders of magnitude higher energy efficiency as compared to ARM-based solutions, highlighting the potential long-term operation in wearable devices. Moreover, this work introduces a novel \gls{sswce} loss function for wearable seizure detection systems. By utilizing subject-specific models and the \gls{sswce} loss function, we achieved an average specificity of $99.74\%$, and detect $92.00\%$ of the seizures with a  false positive rate of $1.18$ FP/h across all subjects in the novel PEDESITE dataset. The results demonstrate the effectiveness of the proposed methodology in balancing sensitivity and specificity for personalized seizure detection. While further reducing the FP/h is needed for clinical implementations, the proposed framework also offers data reduction: the human experts do not need to review the entire recording, but only the epochs automatically detected. This research highlights the importance of personalized model design for wearable seizure detection devices, balancing sensitivity and specificity. Our findings contribute to advancing wearable EEG-based seizure detection systems, potentially improving therapeutic outcomes for individuals with epilepsy. 
\section*{Acknowledgment}
\vspace{-0.2cm}
This project was supported by the Swiss National Science Foundation (Project PEDESITE) under grant agreement 193813 and by the EU project BONSAPPS (g.a. 101015848. This work was also supported by the ETH Future Computing Laboratory (EFCL).
\bibliographystyle{IEEEtran}
\bibliography{bib}
\end{document}